\documentstyle[prl,aps,epsf,graphicx]{revtex}

\newcommand{\beq}{\begin{equation}}
\newcommand{\beqn}{\begin{eqnarray}}
\newcommand{\eeq}{\end{equation}}
\newcommand{\eeqn}{\end{eqnarray}}
\newcommand{\vp}{\varphi}
\newcommand{\dvp}{\delta\phi}
\newcommand{\ts}{\textstyle}
\newcommand{\rd}{\displaystyle{\cdot}}
\def\lb{\label}
\def\sec{\section}

\begin{document}
\draft
\twocolumn[\hsize\textwidth\columnwidth\hsize\csname
@twocolumnfalse\endcsname

\preprint{}

\title{Restoring the sting to metric preheating}

\author{Bruce A. Bassett$^{\dagger\ddagger}$, Christopher
Gordon$^{\ddagger}$, Roy Maartens$^{\ddagger}$, and
David I. Kaiser$^{*}$ }
\address{$\dagger$ Department of Theoretical Physics,
Oxford University, Oxford~OX1~3NP, England}
\address{$\ddagger$ Relativity and Cosmology Group, Division
of Mathematics and Statistics, \\ Portsmouth University,
Portsmouth~PO1~2EG, England}
\address{$^*$  Lyman Laboratory of Physics, Harvard University,
Cambridge, MA~02138 USA}
\date{\today}
\maketitle

\begin{abstract}
The relative growth of field and metric perturbations during
preheating is sensitive to initial conditions set in the preceding
inflationary phase. Recent work suggests this may protect
super-Hubble metric perturbations from resonant amplification
during preheating. We show that this possibility is fragile and
sensitive to the specific form of the interactions
between the inflaton and other fields. The suppression is
naturally absent in two classes of preheating in which either
(1)~the vacua of the non-inflaton fields during inflation are
deformed away from the origin, or (2)~the effective masses of
non-inflaton fields during inflation are small but during
preheating are large. Unlike the simple toy model of a $g^2 \phi^2
\chi^2$ coupling, most realistic particle physics models contain
these other features.  Moreover, they generically lead to both
adiabatic and isocurvature modes and non-Gaussian scars on
super-Hubble scales. Large-scale coherent magnetic fields may also
appear naturally.
\end{abstract}
\pacs{98.80.Cq  \hspace*{2.2cm}hep-ph/9909482}
]
\sec{Introduction}

Standard inflationary models must end with a phase of reheating
during which the inflaton, $\phi$, transfers its energy to other
fields. Reheating itself may begin with a violently nonequilibrium
``preheating" era, when coherent inflaton oscillations lead to
resonant particle production (see \cite{KLS} and refs. therein).
Until recently, preheating studies implicitly assumed that
preheating proceeds without affecting the spacetime metric. In
particular, causality was thought to be a ``silver bullet,"
ensuring that on cosmologically relevant scales, the non-adiabatic
effects of preheating could be ignored.

In fact, exciting, super-Hubble effects are possible during
preheating, and metric perturbations may be resonantly amplified
on all length scales \cite{met,mm,BV}. Causality is not violated
precisely because of the huge coherence scale of the inflaton
immediately after inflation \cite{met} (see also \cite{FB}).
Strong preheating (with resonance parameter $q \gg 1$; see
\cite{KLS,met} for overviews and notation) typically leads to
resonant amplification of scalar metric perturbation modes
$\Phi_k$, including those on super-Hubble scales (i.e., $k/aH \ll
1$, where $a$ is the scale factor and $H$ the Hubble rate). One of
our aims is to answer the question ``how typical is {\em
typical}?"

The answer is crucial since preheating can lead to distortions in the
anisotropies in the cosmic microwave background ({\sc cmb}). Observational
limits rule out those models that produce unbridled nonlinear
growth, but models which pass the metric preheating test on {\sc cobe}
scales may nevertheless leave a non-adiabatic signature of
preheating in the {\sc cmb}. Hence one can no longer universally
avoid consideration of reheating when analyzing inflationary
predictions for cosmology, even if the final effect of reheating in
some particular models is small.

In this vein, it has been argued recently \cite{JS,I} that metric
perturbations on super-Hubble scales are in fact immune to metric
preheating in  the archetypal 2-field potential typically used in
earlier studies \cite{KLS,met}. The claim  arises  because the
initial value of the fluctuations in the created bosonic field
$\chi$ at the start of preheating is much smaller than that used
in \cite{met}. The basic argument is as follows. For the coupling
${1\over2} g^2 \phi^2\chi^2$, strong preheating typically requires
$q \equiv g^2\phi^2/m^2 \gg 1$ (exceptions exist in which $q$ is
small but metric preheating is strong \cite{BV}). This increases
the effective $\chi$ mass relative to the Hubble rate during
inflation, $m_{\chi,{\rm eff}} \sim g\phi \gg H\sim m$, where $m$
is the inflaton mass. This leads to an exponential suppression
$\propto a^{-3/2}$ of both $\chi$ and $\delta\chi_k$ during
inflation; hence these fields would have values at the start of
preheating around $\sim 10^{-36}$ smaller than those used in all
previous simulations. This would stifle any growth in the
small-$k$ modes of $\Phi$ until late times. Initial conditions for
large-$k$ modes, in contrast, are claimed to be unaffected, so
that they would grow nonlinear first. Their resulting backreaction
would then end the resonance before any interesting effects occur
on cosmologically significant scales \cite{JS,I}. Irrespective of
super-Hubble behavior, we note that non-perturbative
metric-preheating effects are vital on
smaller scales \cite{JS,nonlin}, and this in itself is a {\em
major} departure from the old theory
that neglects metric perturbations in preheating.
Metric preheating leads to interesting possibilities, such as
significant primordial black hole formation \cite{met} (see also
\cite{nonlin,JS}).

Returning to super-Hubble scales, $k/aH \ll 1$, we will show that
the above suppression mechanism is {\em highly } sensitive to the
particular form of interaction Lagrangian, while metric preheating
is not. Indeed, the suppression of $\chi$ and $\delta \chi_k$ at
the start of preheating argued for in \cite{JS,I} is {\em absent}
for models in either of the following two classes:

\newpage
\hrule
\vspace*{0.2cm}

{\bf Class I} - Models in which the vacuum expectation value ({\sc
vev}) of $\chi$ is nonzero during inflation.

{\bf Class II} - Models in which the $\chi$ effective mass is
small during inflation but undergoes a transition and becomes
large during preheating.

\vspace*{0.2cm}
\hrule
\vspace*{0.2cm}

Since these possibilities arise naturally in a variety of realistic
particle physics models, we conclude that the suppression mechanism
proposed recently \cite{JS,I} is {\em fragile}, i.e. unstable to
small changes in the potential. On the other hand, resonant growth
of super-Hubble metric perturbations in preheating is {\em robust},
since it persists under small changes of the potential.

The fields split into a homogeneous part and fluctuations:
$\phi_I(t,{\bf x})=\varphi_I(t)+\delta\phi_I(t,{\bf x})$. The
background equations are \beq
 H^2 = {\ts{1\over3}}\kappa^2\left[V+
{\ts{1\over2}}\sum\dot{\varphi}_I^2\right]\,,~
\ddot{\varphi}_I + 3H \dot{\varphi}_I +
  V_{I} = 0 \,,
  \label{bkgd1'}
\eeq where $\kappa^2 \equiv 8\pi/M_{\rm pl}^{2}$ and $V_{I}
\equiv\partial V/\partial\varphi_I$. The linearized equations of
motion for the Fourier modes of field ($\delta\phi_{Ik}$) and
scalar metric fluctuations ($\Phi_k$) are \beqn &&\left(
\delta\phi_{Ik} \right)^{\rd\rd} + 3H \left(\delta\phi_{Ik}
\right)^{\rd} + (k^2/a^2) \delta \phi_{Ik}=\nonumber\\ &&~~~~~~~{}
-\sum V_{IJ}\delta\phi_{Jk} +4\dot{\varphi}_I\dot{\Phi}_k - 2 V_I
\Phi_k \,, \lb{coupledkg} \\ &&\dot{\Phi}_k + H \Phi_k =
{\ts{1\over2}}\kappa^2 \sum \dot{\varphi}_I\delta\phi_{Ik}\,.
\label{multi1} \eeqn This system is subject to the constraint \beq
\left[\frac{k^2}{a^2} - \ts{1\over2}\kappa^2\sum\dot{\varphi}_I^2
\right]\Phi_k= \left.-{\ts{1\over2}}\kappa^2\sum\dot{\vp}_I^2(
\dvp_{Ik}/\dot{\vp}_I)^{\rd}\right.\,, \lb{con} \eeq which we use
to check the accuracy of our numerical integrations of Eqs.
(\ref{coupledkg}) and (\ref{multi1}) and to set $\Phi_k$ initial
conditions.

We envisage a model with the inflaton $\phi_1\equiv\phi=
\varphi(t) + \delta\phi(t,{\bf x})$ coupled to a massless scalar
field $\phi_2\equiv\chi = X(t) + \delta\chi(t,{\bf x})$ (assumed
to be in its vacuum state near the end of inflation). This
schematically represents the particle content of the inflationary
and preheating eras. More realistic models should consider the
gauge group, non-minimal coupling and fermionic effects, and of
course an accurate phenomenology of metric preheating must begin
to study these issues \cite{j}. However, since we are interested
only in essential conceptual points, this simple picture will
suffice for now.

\section{Super-Hubble metric preheating}

The often-used interaction term ${1\over2}g^2 \phi^2 \chi^2$ is
not the only coupling appropriate to preheating, but is rather one
simple example for which resonance occurs. As we show below,
additional couplings linear in $\chi$, as well as quadratic
couplings in which $g^2 < 0$, provide a mechanism for escaping the
inflationary suppression claimed in \cite{JS,I}. Essentially,
these alternatives produce a nonzero attractor $X \neq 0$, to
which inflation drives the $\chi$ field, so that the initial
values of $X$ and $\delta\chi_{k\sim 0}$ at preheating are not
suppressed. These possibilities  are incorporated in the effective
potential
\begin{eqnarray}
V &=& {\ts{1\over2}}m^2\phi^2+{\ts{1\over4}}\lambda\phi^4
+{\ts{1\over4}}{\lambda}_\chi\left(\chi^2-\sigma^2\right)^2
\nonumber\\
&&{}+ {\ts{1\over2}}\epsilon g^2\phi^2\chi^2
 + \tilde{g}^2\kappa^{n-3}\phi^n \chi\,,
\lb{pot}
\end{eqnarray}
for the constants $\epsilon=\pm1$ and $n=2,3$. The $\lambda$,
$\lambda_\chi$ terms ensure that $V$ is bounded from below. The
various terms in this potential are phenomenologically
well-motivated:

\noindent$\bullet$ In theories where supersymmetry ({\sc susy})
is softly broken, the potential will only acquire logarithmic
radiative corrections and the suppression may apply. However, in
realistic models with gravity, {\sc susy} is replaced by
supergravity ({\sc sugra}), and {\sc sugra} models often contain
couplings of the form $\phi^n\chi$, $n = 2,3$ \cite{subir}.
\\
$\bullet$ Even if $\tilde{g} = 0$ initially, if the $\chi$ field
exhibits symmetry-breaking ($\sigma \neq 0$), shifting the field
$\chi \rightarrow\chi - \sigma$ generates the linear term $\sigma
g^2\phi^2\chi$ via the quadratic coupling. The possible importance
of symmetry breaking of this sort has long been noted \cite{KLS94}
for its role in generating single-body inflaton decays and hence
complete inflaton decay. If we choose $\sigma$ to correspond to the
grand unified theory ({\sc gut}) scale, then $\sigma/M_{\rm pl}\sim
\tilde{g}^2/g^2 \sim 10^{-3}$.
\\
$\bullet$ Negative coupling instability ({\sc nci}) models ($\epsilon =
-1$) are dominated by the coupling $-g^2\phi^2\chi^2$,
and the $\chi$ field is driven to a nonzero {\sc vev} during inflation
\cite{GPR}.
\\
$\bullet$ A fermionic coupling $h \overline{\psi} \chi \psi$,
would lead to a driving term  $h \langle\overline{\psi}
\psi\rangle$ in the $\chi$ equation of motion. This would have a
similar effect of giving a nonzero {\sc vev} for $\chi$.

\subsection{Class I: unsuppressed initial conditions}

We now present analytical arguments (assuming for simplicity that
$\sigma = 0$) to show that the new couplings avoid the claimed
suppression of super-Hubble $\chi$ fluctuations. By Eq.
(\ref{bkgd1'}), the background $X$ field obeys \beq \ddot{X} +
3H\dot{X} + \epsilon g^2\varphi^2 X + \lambda_{\chi} X^3 =
-\tilde{g}^2\kappa^{n-3}\varphi^n\,, \lb{chi} \eeq and by Eq.
(\ref{coupledkg}), its large-scale fluctuations satisfy
\begin{eqnarray}
&&(\delta{\chi}_k)^{\rd\rd} + 3H(\delta{\chi}_k)^{\rd} + [\epsilon
g^2\varphi^2+3\lambda_\chi X^2]\delta \chi_k =\nonumber \\
&&~{}4\dot{X}\dot{\Phi}_k+
[(2-n)\tilde{g}^2\kappa^{n-3}\vp^{n-1}+2\lambda_\chi
X^3/\varphi]\delta\phi_k\,, \lb{deltachi}
\end{eqnarray}
where we used the slow-roll relation $\Phi\approx-\delta\phi/\vp$.
We now consider the two separate cases with similar results:

\underline{\em Case 1: $\epsilon = 1, \tilde{g} > 0$,
$\lambda_\chi$ negligible}

Using the fact that $\vp,H\approx$ constant during inflation, we
see that while the solution of the homogeneous part of Eq.
(\ref{chi}) decays rapidly towards zero as $a^{-3/2}$, the
particular solution arising from the inhomogeneous term is
approximately constant. It follows that $\chi$ emerges at the end
of inflation ($t=t_0$) with background part \beq X(t_0) \approx
-(\tilde{g}/g)^2\kappa^{n-3}[\varphi(t_0)]^{n-2}\,, \lb{chipart}
\eeq where $\varphi(t_0)\approx 0.3M_{\rm pl}$. Similarly, the
fluctuations also have a nontransient solution. For $n=2$, we need
to include  the small term $\dot{X}\dot{\Phi}_k$, which is not
straightforward to evaluate, but for $n=3$ we can neglect this
term, and Eq. (\ref{deltachi}) implies
\begin{equation}
\delta\chi_{k}(t_0) \approx -(\tilde{g}/g)^2\delta\phi_k(t_0)\,.
\lb{deltachipart}
\end{equation}
Thus the super-Hubble $\chi$ fluctuations emerge from inflation
unsuppressed, though smaller than the inflaton fluctuations by a
factor $(\tilde{g}/g)^2$.

\underline{\em Case 2: $\epsilon = -1, \tilde{g} = 0$}

{\sc nci} coupling gives rise to a non-zero {\sc vev}, since Eq.
(\ref{chi}) has an attractor solution ($\dot{X}\rightarrow0$), and
then using this solution, the fluctuations governed by Eq.
(\ref{deltachi}) are also seen to have an approximate attractor
solution: \beq X(t_0) \approx (g/\sqrt{\lambda_{\chi}}\,)
\varphi(t_0)\,,~
\delta\chi_k(t_0)\approx(g/\sqrt{\lambda_\chi}\,)\delta\phi_k(t_0)\,.
\label{nci}\eeq

In both cases, for consistency, inflation should be dominated by
the ${1\over2}m^2\phi^2$ term in the potential, and the
super-Hubble fluctuations should be dominated by adiabatic
inflaton fluctuations. The equations for $\varphi$ and
$\delta\phi$ show that this will be secured if $\lambda$ is
negligible and $|\epsilon
g^2X^2+n\tilde{g}^2\kappa^{n-3}\varphi^{n-2} X|\ll m^2$, given
Eqs. (\ref{chipart})--(\ref{nci}). In summary, our analytical
arguments show that by the end of inflation, the $\chi$ field and
its super-Hubble fluctuations are not negligibly small; the linear
couplings ($\tilde{g}>0$, $\epsilon=1$) and the negative quadratic
coupling ($\epsilon=-1$, $\tilde{g}=0$) each provide a mechanism
to evade the super-Hubble suppression of $\chi$ fluctuations.

\subsection{Class I: numerical simulations}

In order to confirm and extend the analytical arguments above, we
performed numerical simulations in one Class I model, with
$\tilde{g}^2\phi^3\chi$ coupling ($\epsilon=1$, $\lambda$ and
$\lambda_\chi$ negligible). To avoid subtleties associated with
matching inflation to preheating, we numerically integrated Eqs.
(\ref{bkgd1'})--(\ref{multi1})
{\em starting deep inside the inflationary era.} Our primary
interest is in cosmologically relevant scales, so we follow the
evolution of a scale that crosses the Hubble radius at $t=t_{\rm
in}$, about 55 $e$-folds before
the start of preheating at $t=t_0$.

The slow-roll approximation gives ${\cal N}=\kappa^2(\vp_{\rm
in}^2-\vp_0^2)/4$ for the number of $e$-folds before the end of
inflation, so we choose $\vp_{\rm in}=3M_{\rm pl}$ to get ${\cal
N}\approx55$. For the background $\chi$ field we use the
approximate dominant solution in Eq. (\ref{chipart}) and take
$X_{\rm in}=-(\tilde g/g)^2\varphi_{\rm in}$. We follow \cite{JS}
and take the field fluctuations at Hubble-crossing ($k=aH$) as
$k^3|\delta\phi_{Ik}|^2= H^3/(2\omega_{Ik})$ and
$|(\delta\phi_{Ik})^{\rd}|=\omega_{Ik}|\delta\phi_{Ik}|$, where
$\omega_{Ik}^2=(k/a)^2+m_I^2$, with $m_\chi=g\vp$. We also take
$\dot{X}_{\rm in}=\omega_\chi X_{\rm in}$. The initial metric
perturbation $(\Phi_{k})_{\rm in}$ is then fixed by Eq.
(\ref{con}). The comoving wavenumber is $k\approx ma(t_0)e^{-{\cal
N}}\kappa\vp_{\rm in}/\sqrt{6}$. We also take $\tilde{g}/g\leq
10^{-2}$, with $g=\sqrt{4\pi/3}\times10^{-3}$ and $m=10^{-6}M_{\rm
pl}$. This yields a resonance parameter $q = 3.8 \times 10^{5}$
which is used for all  our simulations here.

As well as tracking a scale that crosses the Hubble radius at
$t_{\rm in}$, we consider scales that are
within the Hubble radius at $t_{0}$, i.e.,
{\em at the start of preheating}, with $k/a_0H_0>1$.
Although fluctuations on these small scales are
cosmologically insignificant, we need to compare their evolution
with those on very large scales, since this has a bearing on the
question of backreaction. The initial amplitude is given by
$a_{0}^3|\delta\phi_{Ik}|^2=1/(2\omega_{Ik})$, and for $k/a_{0}\gg
g\varphi_{0}\gg m$ we find that $|\delta\phi_{Ik}(t_{0})|=1/
(a_{0}\sqrt{2k})$.

The numerical results summarized in Fig. 1 confirm the analytical
discussion above. The field and metric fluctuations on
cosmological scales are resonantly amplified as expected.
\begin{figure}[h]
\epsfxsize=5.4in
\includegraphics[trim=0 -5 0 0]{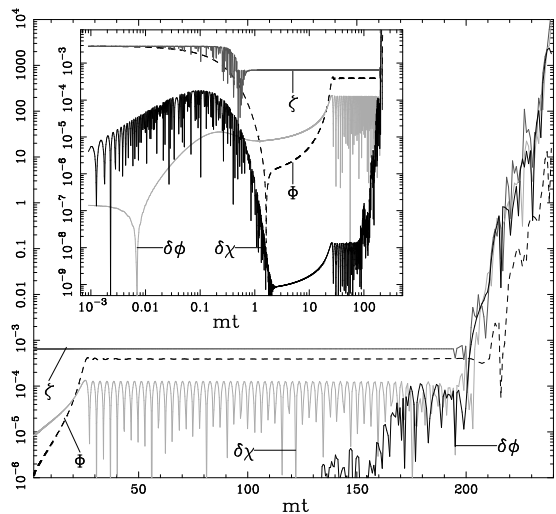}
\caption{ Growth of $|k^{3/2}\Phi_k|$, $|k^{3/2}\zeta_k|$,
$|k^{3/2}\delta\phi_k|/M_{\rm pl}$ and $|k^{3/2}\delta\chi_k|/M_{\rm pl}$ with
$\tilde{g}/g = 10^{-2}$, $k/ma_0 \sim 10^{-23}$ and $q = 3.8
\times 10^5$. {\bf Inset:} Including the detailed evolution during
inflation.} \label{fig1}
\end{figure}
 The
curvature perturbation $\zeta =\Phi-H(\dot{\Phi}+H\Phi)/\dot{H}$,
which
would remain constant in standard reheating on super-Hubble
scales in adiabatic models,
instead shows violently non-adiabatic growth.
This resonant amplification will be terminated by backreaction
effects, which are governed by the growth of the variance
$\langle\chi^2\rangle\propto\int dk\,k^2
|\delta\chi_k|^2$ (suitably renormalized and regularized
\cite{j2}). Resonant growth on small scales will reinforce the
backreaction, since the $k^2$ factor will weight the sub-Hubble
contribution more strongly. Our simulations indicate that for
the chosen value of $q$, resonance occurs for sub-Hubble scales with
$1\leq k/a_0H_0<100$ at the start of preheating,
which occurs at $mt_0\sim 20$.
In Fig. 2 we plot the fluctuations for
a mode with $k/a_0H_0=10$. In addition to the resonance,
this shows that nonlinear growth
in the sub-Hubble mode occurs {\em before} that of the
super-Hubble mode of Fig. 1.
Nonlinear growth of the super-Hubble modes
may therefore be prevented,
but since these modes begin to grow resonantly soon
after the sub-Hubble modes ($\Delta mt\sim 20$), we can
expect some preheating growth in the power spectrum on
cosmological scales.
For other values of $q$ we expect that super-Hubble modes may grow
the quickest, as explicitly occurs in some models which can be
studied analytically \cite{BV}.  The study of backreaction
(including {\em both} gravitational and matter-field
contributions), and of
the preheating imprint on the power spectrum, is currently in
progress.
\begin{figure}[h]
\epsfxsize=5.4in
\includegraphics[trim=0 -5 0 0]{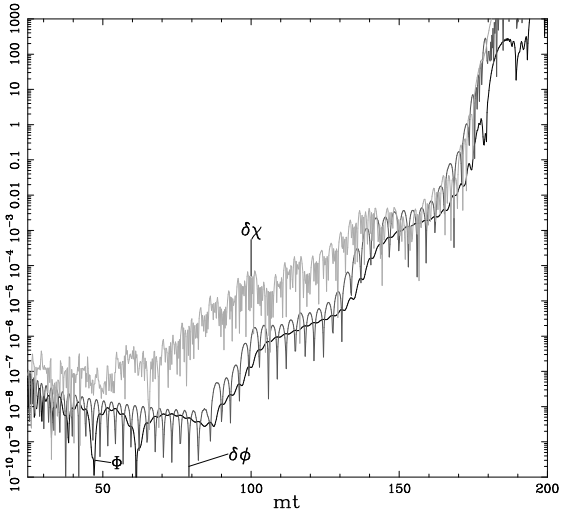}
\caption{ As in Fig. 1, but for a scale that is within the Hubble
radius at the {\em start of preheating} ($mt_0\sim 20$), with
$k/a_0H_0 =10$. } \label{fig2}
\end{figure}

An indication of how the strength of the super-Hubble
resonance in $\Phi_k$ is affected by changes in
the coupling strength $\tilde g/g$ is given in Fig. 3. Here we
have plotted the time $t_{\rm nl}$ when the metric and field
fluctuations  grow to be nonlinear, i.e., $|k^{3/2}\Phi_k(t_{\rm
nl})|\sim1$, $|k^{3/2}\delta\phi_{Ik}|\sim M_{\rm pl}$. The results show how
$t_{\rm nl}$ increases  in response to the suppression of initial
conditions that occurs as $\tilde{g}$ is decreased. Note that
synchronization occurs:  all fluctuations share roughly the same
$t_{\rm nl}$ values
so that we expect $\langle\Phi^2\rangle \sim \langle\phi^2\rangle / M_{\rm pl}^2,
\langle\chi^2\rangle / M_{\rm pl}^2$. The importance of metric perturbations in
determining backreaction has been independently noted in recent work
\cite{nonlin}.
\begin{figure}
\epsfxsize=5.4in
\includegraphics[trim=0 -5 0 0]{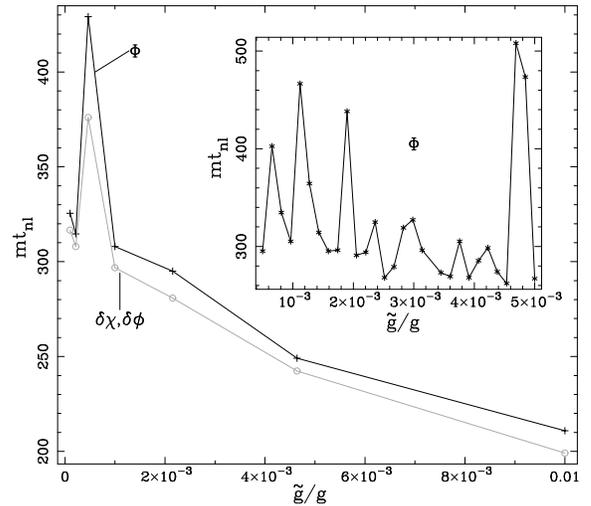}
\caption{ The time to nonlinearity for super-Hubble perturbations
as $\tilde{g}/g$ increases from $10^{-4}$ to $10^{-2}$ ($q = 3.8
\times 10^5$). On average, $t_{\rm nl}$ decreases rapidly as
$\tilde{g}/g$ increases. {\bf Inset:} a zoom with $5\times
10^{-4}\leq\tilde{g}/g\leq 5\times 10^{-3}$. } \label{fig3}
\end{figure}

\subsection{Class II models}

In Class II models, the $\chi$ effective mass is simply very small
during inflation but then becomes large at preheating. This occurs
naturally in various models:

\noindent$\bullet$ Globally {\sc susy} hybrid models based on the
superpotential $W=\alpha S \overline{\varphi} \varphi - \mu^2 S$,
where the singlet $S$ plays the role of the inflaton. The
corresponding unbroken potential is $V = \alpha^2
|S^2|(|\varphi|^2 + |\overline{\varphi}|^2) + |\alpha \varphi
\overline{\varphi} - \mu^2|^2$, together with $D$-terms which
vanish along the flat direction $|\varphi| =
|\overline{\varphi}^*|$.  For $S \gg \mu/\sqrt{\alpha}$, inflation
occurs with the minimum of the potential at $\langle
\varphi\rangle = \langle \overline{\varphi} \rangle= 0$. However,
for $S \leq \mu/\sqrt{\alpha}$, $V$  has a new minimum at $\langle
S \rangle = 0, \langle \varphi \rangle = \mu/\sqrt{\alpha}$ and
preheating occurs via oscillations around this minimum \cite{hyb}.
Now let us couple $\chi$ {\em not} to the inflaton $S$, but to the
field $\varphi$ through the term $g^2 \chi^2 |\varphi|^2$. Then
the $\chi$ effective mass $g |\varphi|$ {\em vanishes} during
inflation (up to logarithmic corrections)
-- and hence so does the suppression mechanism of \cite{JS,I}.
The effective mass only departs
strongly from zero once inflation ends and reheating begins,
leading to a huge increase in the value of the resonance parameter
$q$.

\noindent$\bullet$ In models with strong running of coupling
constants, where the beta function is negative, such as occurs in
{\sc qcd}, the theory is asymptotically free and the coupling {\em
increases} at lower energies. Perhaps the strongest examples of
this are based on $S$-type dualities, where the coupling $g^2$ is
very small during inflation but is very large  during reheating,
which occurs in the strongly coupled phase with dual coupling
$\propto 1/g^2 \gg 1$. An example is provided by `dual inflation'
\cite{GB}, where $m_{\chi, \rm eff} \sim g\phi < H$, and $\chi$
fluctuations are similar to those in the inflaton, and not
strongly suppressed. In fact, it is arguable that models of this
sort are needed if preheating is to be viable in non-{\sc susy}
theories, since large $g$ leads to radiative corrections to the
potential which may violate the slow-roll conditions for
inflation.

\section{New Cosmological effects}

Our eventual goal  must  be to calculate physical quantities such
as the power spectrum of $\Phi_k$. Since $P_{\Phi} = k^3
|\Phi_k|^2/2\pi^2$, one might be concerned that these strong
preheating effects at $k \rightarrow 0$ would be made irrelevant
by the $k^3$ phase space factor. Perhaps the easiest way to see
that this is not so is to look at the evolution of $\zeta_k$.
Since $\zeta_k$ is {\em not} conserved for small  $k$ (see Fig.
\ref{fig1}), the standard normalization of the {\sc cmb} spectrum
is increased. This can only take place if the power spectrum of
metric fluctuations is strongly affected as $k \rightarrow 0$.
This is understandable since preheating acts only as a non-trivial
transfer function $T(k)$.

Beyond the effects discussed in \cite{met},
metric preheating can lead to a host of interesting new effects.\\
$\bullet$ The growth of
$\zeta_k$ implies amplification of isocurvature modes in unison
with adiabatic scalar modes on super-Hubble scales. Preheating
thus yields the possibility of inducing a post-inflationary
universe with both isocurvature and adiabatic modes on large
scales. If these are uncorrelated and of roughly equal strength,
the corresponding Doppler peaks will tend to cancel
\cite{doppler}. (This mechanism is independent of the one
discussed in \cite{met}, which requires nonlinearity to persist
until decoupling.) However, if the adiabatic and isocurvature
modes are strongly correlated, this would create the possibility
of a ``smoking gun" finger-print of preheating. The challenge
remains to distinguish such correlations from those induced in
hybrid inflation.\\ $\bullet$ Because the metric perturbations can
go nonlinear, whether on sub- or super-Hubble scales, the
corresponding $\chi$ density perturbations $\delta$ typically have
non-Gaussian statistics. This is simply a reflection of the fact
that $-1\leq\delta <\infty$, so that the distribution of necessity
becomes skewed and non-Gaussian. Further, in Class II models,
where $\langle\chi\rangle = 0$ during inflation, $\chi$
perturbations in the energy density will necessarily be
non-Gaussian (chi-squared distributed), even if $\delta\chi_k$ is
Gaussian distributed, since stress-energy components are quadratic
in the fluctuations (see e.g. \cite{nongauss}).  Non-Gaussian
effects are therefore an intrinsic part of many metric preheating
models (particularly those in Class II), and open up a potential
signal for detection in future experiments.\\ $\bullet$ Another
new feature we would like to identify is the breaking of conformal
invariance. Once metric perturbations become large on some scale,
the metric on that scale cannot be thought of as taking the simple
Friedmann-Lemaitre-Robertson-Walker ({\sc flrw}) form, and
conformal invariance is lost. This is particularly important for
the production of  primordial magnetic fields, which are usually
strongly suppressed due to the conformal invariance of the Maxwell
equations in a {\sc flrw} background. The coherent oscillations of
the inflaton during preheating further provide a natural cradle
for producing a primordial seed for the observed large-scale
magnetic fields. A charged inflaton field, with kinetic term
$D_{\mu} \phi (D^{\mu}\phi)^*$, will  couple to electromagnetism
through the gauge covariant derivative $D_{\mu} = \nabla_{\mu} -
ie A_{\mu}$. This  will naturally lead to parametric resonant
amplification of the existing magnetic field, which
could produce
large-scale coherent seed fields on the required super-Hubble
scales without fine-tuning
\cite{pbvm}. (Note that a tiny seed field {\em must} exist
during inflation due to the conformal trace anomaly and one-loop
QED corrections in curved spacetime \cite{mag}.)\\

In conclusion, the suppression discussed in \cite{JS,I} is highly
sensitive to the form of the particle interactions considered;
when couplings are considered which are found in most realistic
particle physics models, the effects of \cite{JS,I} recede.
Instead, in models from either of the two general classes
highlighted here, preheating
can produce a strong amplification of metric perturbations on
cosmologically significant scales. Metric preheating thus allows
us to rule out models in which backreaction effects fail to
prevent
super-Hubble nonlinear growth, and shows that in the surviving
models, there will typically be some signature of preheating
imprinted on the power spectrum. The robustness of the
amplification further demonstrates the need to move towards more
realistic models of preheating in order to develop a realistic
understanding of the predictions of inflation for observational
cosmology.

We thank Kristin Burgess for detailed comments and
Paul Anderson, J\"urgen Baacke, Arjun Berera, Anthony Challinor, Fabio
Finelli, Alan Guth, Karsten Jedamzik, Dimitri Pogosyan, Giuseppe
Pollifrone, Dam Son, Alexei Starobinsky and David Wands for stimulating
discussions. DK receives partial support from NSF grant
PHY-98-02709.


\end{document}